# On the Asymptotic Convergence of the Transient and Steady State Fluctuation Theorems.

*Gary Ayton and Denis J. Evans*

Research School Of Chemistry

Australian National University

Canberra, ACT 0200 Australia

**Abstract**

Non-equilibrium molecular dynamics simulations are used to demonstrate the asymptotic convergence of the *Transient* and *Steady State* forms of the Fluctuation Theorem. In the case of planar Poiseuille flow, we find that the Transient form, valid for all times, converges to the Steady State form on microscopic time scales. Further, we find that the time of convergence for the two Theorems coincides with the time required for satisfaction of the asymptotic Steady State Fluctuation Theorem.

PACS numbers: 05.20.-y, 05.70.Ln, 47.10.+g, 47.40.-n

The Fluctuation Theorem (FT) [1-5] gives the probability ratio of observing fluctuations in a nonequilibrium system in which for N particles studied over a time t, the time averaged entropy production takes on some specified value compared to the negative of that same value. In the long time and large N limit, the Second Law of Thermodynamics is recovered, and for finite systems studied over short times, FT gives a quantitative explanation of the origins of irreversibility in systems with reversible dynamics. Evans and Searles [2-4] employed a Liouville measure and derived a *Transient* FT (TFT) which is exact for all finite times for systems starting at equilibrium and evolving towards a *Steady State* FT (SSFT) as $t \to \infty$. Gallavotti and Cohen [5] derived a steady state FT using the SRB measure and the chaotic hypothesis resulting in an asymptotic formula for the probability ratio. It would seem that if the nonequilibrium steady state is unique, then in the long time limit the Transient and Steady State FT's would converge to the same statement about fluctuations in entropy production. Asymptotic convergence is based on the independence of the final steady state measure with respect to the particular choice of the initial phase point (ie initial transients do not affect steady state averages or statistics). However there has been some recent discussion on this point.

In this Letter we will present convincing non-equilibrium molecular dynamics (NEMD) simulation results that demonstrate the asymptotic convergence of the Transient and Steady State Fluctuation Theorems. We will show that the Transient FT holds for all time, and that after a finite time, the two Theorems converge to the *same* result. Not only does TFT approach the SSTF, but the time of convergence for the two Theorems coincides with the corresponding time of convergence for the asymptotic SSTF itself.

The simulation was exactly as described in [6], and we will only very briefly discuss some details here. We simulated planar Poiseuille flow where an atomic fluid obeying Newton's equations of motion is placed between two heat extracting walls. Heat sinks in the walls remove heat at precisely the rate required to make the total energy of the system a constant of the motion. As in reference [6] we calculated the *Integrated* form of the Transient and Steady State FT's (TIFT and SSIFT respectively), written as

$$p_-(t) / p_+(t) = \left\langle \exp[-3N_W \,\overline{\alpha}(t) t] \right\rangle_+ \equiv \phi(t)\,,$$

where $p_-/p_+$ is the probability of observing an anti-trajectory versus a trajectory, $N_W$ is the number of ergostatting wall particles, $\alpha = -J(\mathbf{\Gamma})VF_e / 2K_W$ is the thermostat multiplier, $K_W = \sum_{i=1}^{N_W} p_i^2 / 2m$ is the kinetic energy of the wall particles, $F_e$ is the external field, the dissipative flux, J is defined, $-J(\mathbf{\Gamma})V \equiv \int d\mathbf{r}\, n(\mathbf{r})u_x(\mathbf{r})$, where $n(\mathbf{r})$ and $u_x(\mathbf{r})$ are the density and flow velocity. The averages, $\langle ... \rangle_+$ denote averages over all trajectory segments for which $\overline{\alpha}(t) = 1/t\int_0^t \alpha(t')dt' > 0$. The notation is identical to reference [6].

Testing the validity of TIFT and SSIFT with NEMD requires a method of generating either a set of transient trajectories or a single long steady state trajectory. In the transient case, two hundred transient nonequilibrium trajectory segments were generated from a microcanonically distributed ensemble of initial phase configurations, $\{\mathbf{\Gamma}_{eq}\}$. Each transient segment was studied for t=10 (the integration time step was $\delta t = 0.001$). This time is sufficiently long that all time averaged properties have converged to their steady state values.

Each transient segment originated from a configuration, $\mathbf{\Gamma}_{eq}$ which was randomly chosen from an equilibrium microcanonical ensemble. It was then simultaneously subjected to an external field ($F_e$=0.032) and initialised as a new transient segment time origin. After following the transient segment for t=10, the trajectory was terminated, and a new equilibrium configuration was selected. This process is shown in Fig. 1a, where the equilibrium microcanonical ensemble is shown by the dotted line, and the initial transient trajectory configuration, $\mathbf{\Gamma}_{eq}$ is designated by a filled circle. A TTIFT can then be tested by examining each of the transient segments at equal time intervals from their respective transient time origins (shown as small circles along the transient trajectories).

It is not possible to generate an exact steady state trajectory. This is because within phase space, the measure of any dissipative nonequilibrium steady state is zero. Therefore the probability of selecting initial phase points that lie exactly on the steady state attractor is zero. We can only *approach* the nonequilibrium steady state. We used an *equilibration method* of approaching the steady state: an arbitrary microcanonical phase point was chosen; an $F_e$=0.032 external field was applied and the system was allowed to equilibrate towards the steady state for a time (t=500). This time is *very much* greater than the decay time of transient time averages (t = 3 - 4). Subsequently a t=5000 "steady state" trajectory was generated. This process is sketched in Fig. 1b where the initial transient segment is shown by the dotted line, and the t=5000 steady state segment is the solid line.

The "steady state" trajectory was decomposed into 2.5 X $10^4$ "steady state" subsegments each of duration t=0.2, (shown as small circles on the steady state trajectory) which could be examined to test the SSIFT with observation times ranging from t=0.2 ( with 2.5 X $10^4$ possible samples for $\overline{\alpha}_+(t_1)$ ), t=0.4 ( with 1.25 X $10^4$ possible samples for $\overline{\alpha}_+(t_2)$ ) to t=5000 (one sample of $\overline{\alpha}_+(t_{5000})$).

Convergence of the TIFT and SSIFT can be tested by exploiting the near exponential decay of $p_-(t)/p_+(t)$ and $\phi(t)$ [see Fig. 3 inset] and calculating $\mathrm{Ln}[p_-(t)/p_+(t)]/\mathrm{Ln}[\phi(t)] = Y(t)$, where $\lim_{t\to\infty} Y(t) = 1$. In Fig. 2 we show $Y_T(t)$ for the transient and $Y_{SS}(t)$ steady state segments. In accord with FT both converge to 1. As expected, $Y_T(t) = 1, \forall t$, but $Y_{SS}(t)$ is only 1 asymptotically (for t > ~4). An alternative test for the asymptotic convergence of the two FT theorems is to examine $\mathrm{Ln}[p_-(t)/p_+(t)]_{SS}/\mathrm{Ln}[p_-(t)/p_+(t)]_T = Z_p(t)$, and $\mathrm{Ln}[\phi(t)]_{SS}/\mathrm{Ln}[\phi(t)]_T = Z_\phi(t)$ where $\lim_{t\to\infty} Z_p(t) = 1$ and $\lim_{t\to\infty} Z_\phi(t) = 1$. In Fig. 3 one clearly sees that both converge to 1 by a time t ~ 4. This test shows that at long times, the *fluctuations* in the transient states converge to the *fluctuations* of the steady states. We note that $Z_p(t)$, $Z_\phi(t)$, and $Y_{SS}(t)$ converge to unity in approximately the same time.

From examining the Transient and Steady State Fluctuation Theorems for a Poiseuille flow system we find strong numerical evidence that, as expected, when the nonequilibrium steady state is unique, the Transient FT asymptotically converges to the Steady State FT at long times. Further, the time scale for this convergence is the same as the time required for the satisfaction of the Steady State FT itself. For our system, convergence of the two Theorems occurs on a microscopic time scale. These results confirm the assumptions of [2-4]. We note that the assumption of a unique nonequilibrium steady state is implicit in all linear and nonlinear response theory. Finally, we have recently shown that for stochastic systems the Transient and Steady State FT's show an analogous asymptotic convergence at long times [7].

## References

[1] D. J. Evans, E. G. D. Cohen and G. P. Morriss, Phys. Rev. Lett., **71**, 2401 (1993).

[2] D. J. Evans and D. J. Searles, Phys. Rev. E, **50,** 1645 (1994).

[3] D. J. Evans and D. J. Searles, Phys. Rev. E, **52**, 5839 (1995).

[4] D. J. Evans and D. J. Searles, Phys. Rev. E, **53**, 5808 (1996).

[5] G. Gallavotti and E. G. D. Cohen, Phys. Rev. Lett., **74**, 2694 (1995);
G. Gallavotti and E. G. D. Cohen, J. Stat. Phys,. **80**, 931 (1995).

[6] Gary Ayton, Denis J. Evans, and D. J. Searles, "A Localised Fluctuation Theorem", submitted to PRL, and at http://xxx.lanl.gov/abs/cond-mat/9901256.

[7] D. J. Evans and D. J. Searles, Phys. Rev. E, "The Fluctuation Theorem for Stochastic Systems", submitted to Phys. Rev. E, and at http://xxx.lanl.gov/abs/cond-mat/9901258.

**Figures**

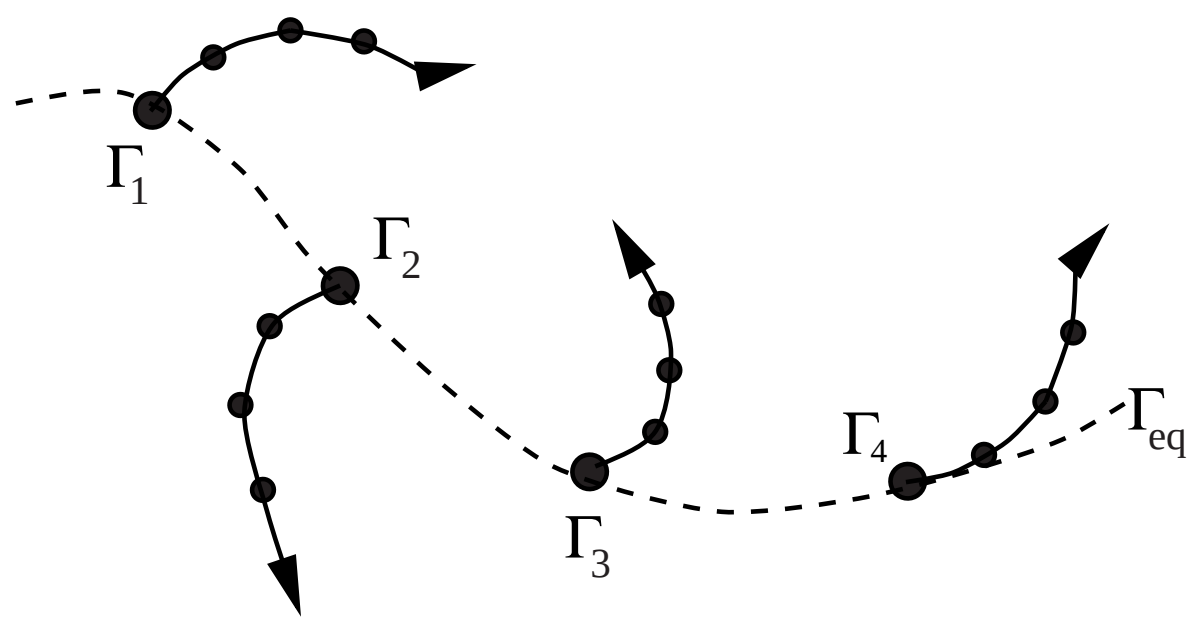

Figure 1a. A sketch of method used to generate transient trajectories. The dotted line refers to an equilibrium microcanonical ensemble. Transient trajectories $\{\mathbf{\Gamma}_i\}$, i=1,200 (solid lines) are studied for a total time of t=10, but averages are accumulated every t=0.2 in order to construct a TIFT.

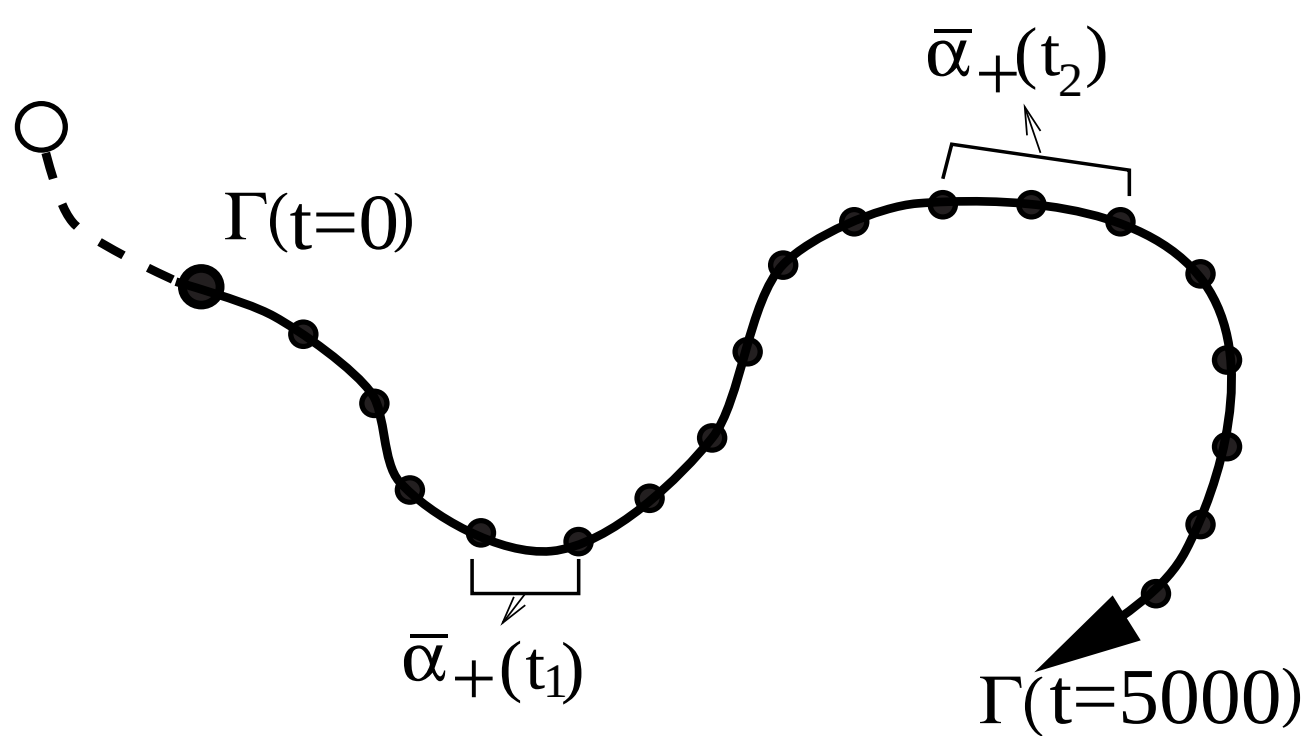

Figure 1b. A sketch of method used to generate steady state trajectories. The dotted line refers to the initial transient trajectory with $F_e$ = 0.032. At the end of the t = 500 equilibration run, the time origin for the steady state trajectory $\mathbf{\Gamma}$ is established (large filled circle), and is studied for t = 5000. Time segments of t = 0.2 were used to construct a SSIFT for the corresponding field strength.

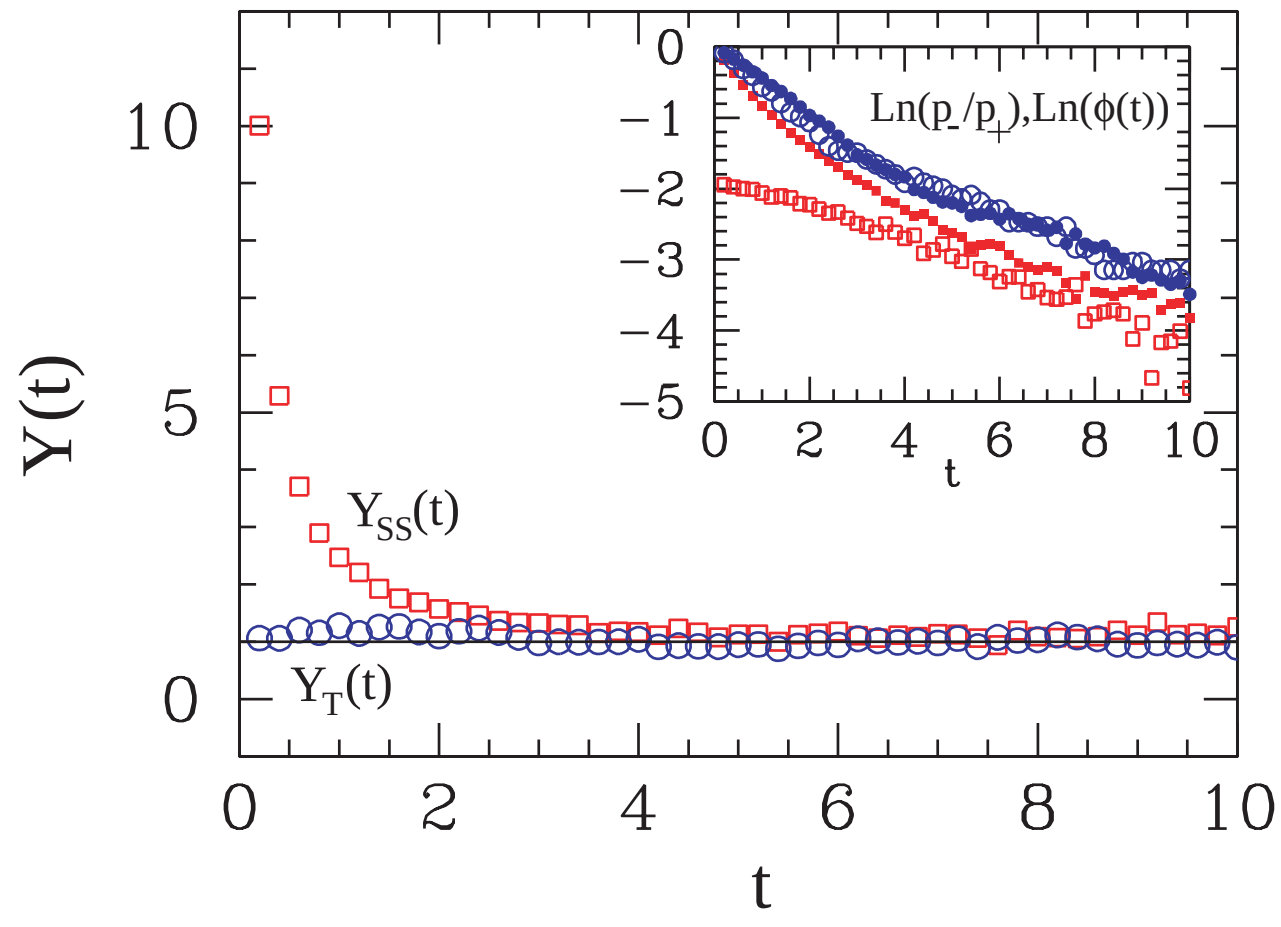

Figure 2. A plot of $Y_T(t)$ for Transient (circles) and $Y_{SS}(t)$ for Steady States (squares) IFT for $F_e = 0.032$. The inset shows the corresponding $\mathrm{Ln}(p_(t)/p_+(t))$ (open symbol) and $\mathrm{Ln}(\phi(t))$ (shaded symbol).

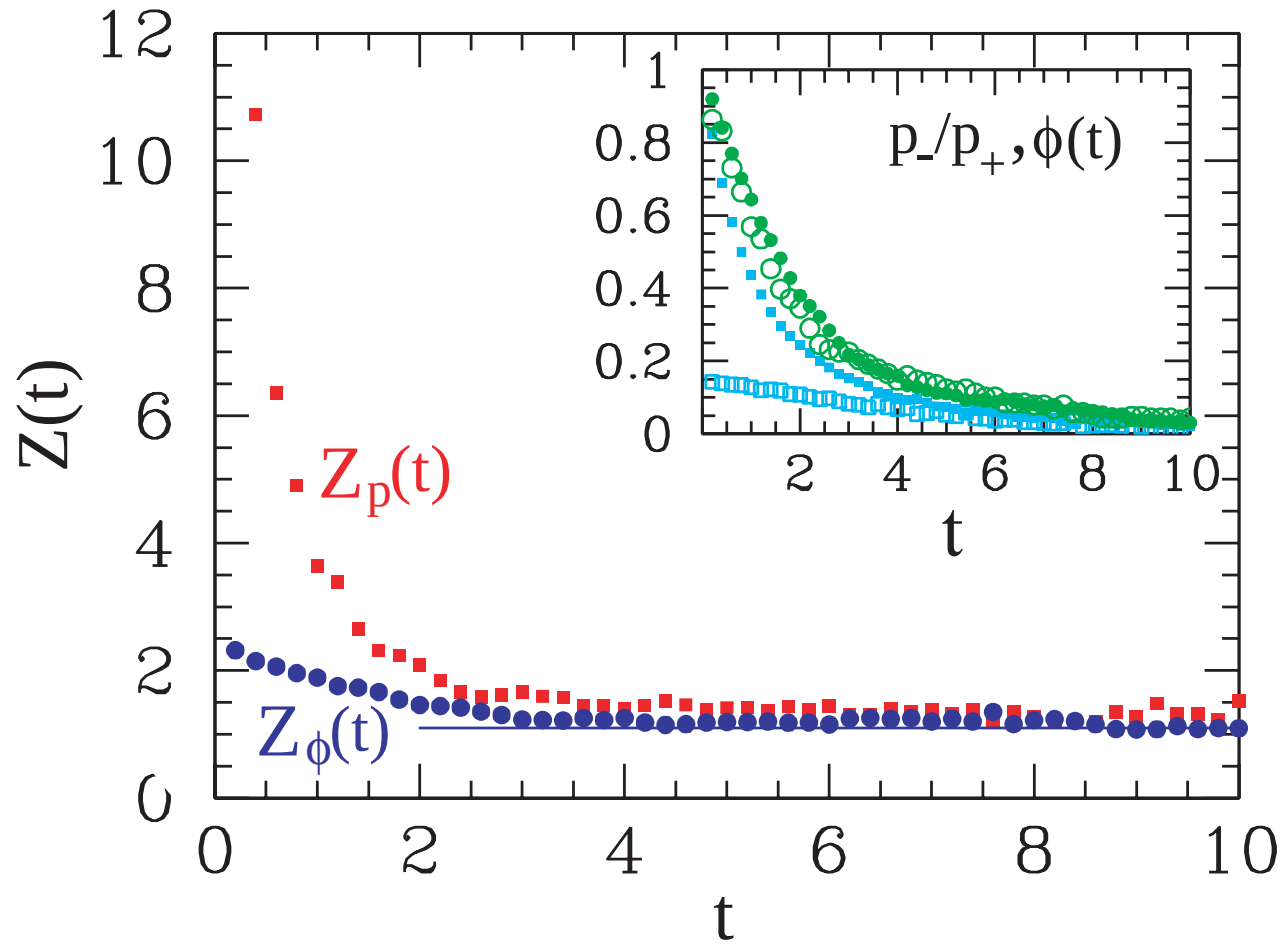

Figure 3. A plot of $Z_p(t)$ (squares) and $Z_\phi(t)$ (circles) for IFT with $F_e = 0.032$. The inset shows the corresponding $p_(t)/p_+(t)$ (open symbol) and $\phi(t)$ (shaded symbol).